\documentclass[twoside,leqno,twocolumn]{article}  
\usepackage{ltexpprt} 

\usepackage{latexsym}
\usepackage{amssymb}  
\usepackage{amsmath}
\usepackage{ifthen}
\usepackage{array}

\newcommand{\defemph}[1]{\textbf{\textsl{#1}}}
\newcommand{\uu}{\mathbf{u}}
\newcommand{\vv}{\mathbf{v}}

\begin{document}

\bibliographystyle{plain}

\title{\Large Positivity Problems for Low-Order Linear 
Recurrence Sequences\thanks{This research was partially supported by
  EPSRC\@. We are also grateful to Matt Daws for considerable
assistance in the initial stages of this work.}}

\author{Jo\"el Ouaknine\thanks{Department of Computer Science,
Oxford University, UK.} \\
\and 
James Worrell\thanks{Department of Computer Science,
Oxford University, UK.}} 
\date{}

\maketitle

\begin{abstract} \small\baselineskip=9pt
We consider two decision problems for linear recurrence sequences
(LRS) over the integers, namely the \emph{Positivity Problem} (are all
terms of a given LRS positive\@?) and the \emph{Ultimate Positivity
  Problem} (are all but finitely many terms of a given LRS
positive?). We show decidability of both problems for LRS of order $5$
or less, with complexity in the Counting Hierarchy for Positivity, and
in polynomial time for Ultimate Positivity. Moreover, we show by way
of hardness that extending the decidability of either problem to LRS
of order $6$ would entail major breakthroughs in analytic number theory,
more precisely in the field of Diophantine approximation of
transcendental numbers.
\end{abstract}

\section{Introduction}

A (real) \defemph{linear recurrence sequence (LRS)} is an infinite
sequence $\uu = \langle u_0, u_1, u_2, \ldots \rangle$ of real numbers
having the following property: there exist constants $a_1, a_2,
\ldots, a_k$ (with $a_k \neq 0$) such that, for all $n \geq 0$,
\begin{equation}
\label{rec-rel}
u_{n+k} = a_1 u_{n+k-1} + a_2 u_{n+k-2} + \ldots + a_k u_{n} \, .
\end{equation} 
If the initial values $u_0, \ldots, u_{k-1}$ of the sequence are
provided, the recurrence relation defines the rest of the sequence
uniquely. Such a sequence is said to have \defemph{order}
$k$.\footnote{Some authors define the order of an LRS as the
  \emph{least} $k$ such that the LRS obeys such a recurrence
  relation. The definition we have chosen allows for a simpler
  presentation of our results and is algorithmically more convenient.}

The best-known example of an LRS was given by Leo\-nar\-do of Pisa in
the 12th century: the Fibonacci sequence $\langle 0,
1, 1, 2, 3, 5, 8, 13, \ldots \rangle$, which satisfies the
recurrence relation $u_{n+2} = u_{n+1} + u_n$. Leonardo of Pisa
introduced this sequence as a means to model the growth of an
idealised population of rabbits. Not only has the Fibonacci sequence
been extensively studied since, but LRS now form a vast subject in
their own right, with numerous applications in mathematics and other
sciences. A deep and extensive treatise on the mathematical aspects of
recurrence sequences is the recent monograph of Everest \emph{et
  al.}~\cite{BOOK}.

In this paper, we focus on two key \emph{decision problems} for LRS
over the integers (or equivalently, for our purposes, the rationals):

\begin{itemize}

\item The \defemph{Positivity Problem}: given an LRS $\uu$, are all
  terms of $\uu$ positive?

\item The \defemph{Ultimate Positivity Problem}: given an LRS $\uu$,
  are all but finitely many terms of $\uu$ positive?\footnote{Note
    that both problems come in two natural flavours, according to
    whether strict or non-strict positivity is required. This paper
    focusses on the non-strict version, but alternatives and
    extensions (including strictness) are discussed in
    Section~\ref{conclusion}.}

\end{itemize}

These problems (and assorted variants) have applications in a wide
array of scientific areas, such as theoretical biology (analysis of
L-systems, population dynamics)~\cite{LR76}, economics (stability of
supply-and-demand equilibria in cyclical markets,
multiplier-accelerator models)~\cite{Bau70}, software verification
(termination of linear
programs)~\cite{PR04,Tiw04,Bra06,CPR11,BIK12,BGM12}, probabilistic
model checking (reachability and approximation in Markov chains,
stochastic logics)~\cite{BRS06,AAG12}, quantum computing (threshold
problems for quantum automata)~\cite{BJK05,DJK05}, discrete linear
dynamical systems (reachability and invariance
problems)~\cite{KL86,TV11,Ben13,COW13}, as well as combinatorics,
formal languages, statistical physics, generating functions, etc. For
example, as discussed in~\cite{Liu10}, terms of an LRS usually have
combinatorial significance only if they are positive. Likewise, an LRS
modelling population size is biologically meaningful only if it is
uniformly positive.

Both Positivity and Ultimate Positivity bear some relationship to the
well-known \emph{Skolem Problem}: does a given LRS have a zero? The
decidability of the Skolem Problem is generally considered to have
been open since the 1930s (notwithstanding the fact that algorithmic
decision issues had not at the time acquired the importance that they
have today---see~\cite{TUCS05} for a discussion on this subject; see
also~\cite{Tao07} and \cite{Lip09}, in which this state of
affairs---the enduring openness of decidability for the Skolem
Problem---is described as ``faintly outrageous'' by Tao and a
``mathematical embarrassment'' by Lipton). A breakthrough occurred in
the mid-1980s, when Mignotte \emph{et al.}~\cite{MST84} and
Vereshchagin~\cite{Ver85} independently showed decidability for real
algebraic LRS of order $4$ or less. These deep results make essential
use of Baker's theorem on linear forms in logarithms (which earned
Baker the Fields medal in 1970), as well as a $p$-adic analogue of
Baker's theorem due to van der Poorten. Unfortunately, little progress
on that front has since been recorded.\footnote{A proof of
  decidability of the Skolem Problem for LRS of order 5 was announced
  in~\cite{TUCS05}. However, as pointed out in~\cite{OW12}, the proof
  seems to have a serious gap.

It is worth remarking, on the other hand, that whether an integer LRS has
\emph{infinitely} many zeros is known to be decidable at all
orders~\cite{BM76}.} The Skolem Problem is known to be
NP-hard if the order is unrestricted~\cite{BP02}.

It is considered folklore that the decidability of Positivity would
entail that of the Skolem Problem (see Section~\ref{sec-LRS}), noting
however that the reduction increases the order of LRS
quadratically. Nevertheless, the earliest explicit references in the
literature to the Positivity and Ultimate Positivity Problems that we
have found are from the 1970s (see,
e.g.,~\cite{Soi76,Sal76,BM76}). In~\cite{Soi76}, the Skolem and
Positivity Problems are described as ``very difficult'', whereas
in~\cite{RS94}, the authors assert that the Skolem, Positivity, and
Ultimate Positivity Problems are ``generally conjectured [to be]
decidable''. Positivity and/or Ultimate Positivity are again stated as
open in~\cite{HHH06,BG07,LT09,Liu10,TV11,TPL12}, among others.

Unsurprisingly, progress on the Positivity and Ultimate Positivity
Problems has been fairly slow. In the early 1980s, Burke and Webb
showed that Ultimate Positivity is decidable for LRS of
order~$2$~\cite{BW81}, and nine years later Nagasaka and
Shiue~\cite{NS90} showed the same for LRS of order~$3$ that have
repeated characteristic roots. Much more recently, Halava \emph{et
  al.}\ showed that Positivity is decidable for integer LRS of
order~$2$~\cite{HHH06}, and three years later Laohakosol and
Tangsupphathawat proved that both Positivity and Ultimate Positivity
are decidable for integer LRS of order~$3$~\cite{LT09}.  In 2012, an
article claiming to show decidability of Positivity for LRS of
order~$4$ was published~\cite{TPL12}, with the authors noting being
unable to tackle the case of order~$5$. Unfortunately, as acknowledged
subsequently by the authors themselves~\cite{Lao13}, that paper
contains a major error (the purported proof of Claim~2 on p.141, aimed
at handling the most difficult critical case at order~$4$, is wrong,
and appears not to be fixable without making use of sophisticated
tools from analytic number theory as is done in the present paper).

To the best of our knowledge, no complexity bounds currently exist in
the literature concerning either the Positivity or Ultimate Positivity
Problems, other than coNP-hardness for LRS of unbounded orders which is
inherited from the reduction from the Skolem
Problem~(cf.\ Section~\ref{sec-LRS}).

Our main results are as follows:\footnote{The complexities are given
  as a function of the bit length of standard representations of
  integer LRS of order $k$; for an LRS as defined by
  Equation~(\ref{rec-rel}), this representation consists of the
  $2k$-tuple $(a_1, \ldots, a_k, u_0, \ldots, u_{k-1})$ of integers.

Note also that the Counting-Hierarchy complexity class does not
  require parenthesising since
  $\mathrm{co(NP}^{\mathrm{PP}^{\mathrm{PP}^{\mathrm{PP}}}}\mathrm{)}
  = \mathrm{(coNP)}^{\mathrm{PP}^{\mathrm{PP}^{\mathrm{PP}}}}$.}

\begin{itemize}

\item The Positivity Problem is decidable for integer LRS of order $5$
  or less, with complexity in
  $\mathrm{coNP}^{\mathrm{PP}^{\mathrm{PP}^{\mathrm{PP}}}}$, i.e.,
  within the fourth level of the Counting Hierarchy.

\item The Ultimate Positivity Problem is decidable for integer LRS of
  order $5$ or less in polynomial time.

\item The decidability of either Positivity or Ultimate Positivity for
  integer LRS of order $6$ would entail major breakthroughs in
  analytic number theory (certain open problems in Diophantine
  approximation of transcendental numbers long believed to be hard
  would become solvable)---see Section~\ref{hardness} for precise
  statements.
\end{itemize}

These results, which---absent major advances in number theory---can
essentially be viewed as completing the picture on Positivity problems
for linear recurrence sequences, substantially improve the state of
the art over the last three decades' worth of research on the
subject. Most prior work on Positivity problems that we are aware of
has been confined to the use of linear algebra and elementary
algebraic number theoretic techniques. By contrast, we are deploying
in this paper an eclectic arsenal of deep and sophisticated
mathematical tools from analytic and algebraic number theory,
Diophantine geometry and approximation, and real algebraic geometry,
notably Baker's theorem on linear forms in logarithms, Masser's
results on multiplicative relationships among algebraic numbers,
Kronecker's theorem on simultaneous Diophantine approximation, and
Renegar's work on the fine-grained complexity of the first-order
theory of the reals. These results are summarised in
Section~\ref{tools}. We then present a high-level overview of our
proof strategy---split in two parts---in the first half of
Section~\ref{decidability}. Various extensions and generalisations of
our results, along with avenues for future work, are discussed in
Section~\ref{conclusion}.

\section{Linear Recurrence Sequences}
\label{sec-LRS}

We recall some fundamental properties of linear recurrence sequences.
Results are stated without proof, and we refer the reader
to~\cite{BOOK,TUCS05} for details.

Let $\uu = \langle u_n \rangle_{n=0}^{\infty}$ be an LRS of order $k$
over the reals satisfying the recurrence relation 
\[ u_{n+k} = a_1 u_{n+k-1} + \ldots + a_k u_{n} \, ,
\]
where without loss of generality we
may assume that $a_k \neq 0$.  We denote by $||\uu||$ the bit length
of its representation as a $2k$-tuple of integers, as discussed in the
previous section. The \defemph{characteristic polynomial} of $\uu$ is
\[p(x) = x^n - a_1 x^{n-1} - \ldots - a_{k-1} x - a_k \, .\]
The \defemph{characteristic roots} of $\uu$ are the roots of this
polynomial, and the \defemph{dominant roots} are the roots of maximum
modulus.

The characteristic roots divide naturally into those that are
real and those that are not. As we exclusively deal with LRS over the
reals, the characteristic polynomial has real coefficients and
non-real roots therefore always arise in conjugate pairs. Thus we may
write $\{\rho_1, \ldots, \rho_{\ell}, \gamma_1, \overline{\gamma_1},
\ldots, \gamma_m, \overline{\gamma_m}\}$ to represent the set of
characteristic roots of $\uu$, where each $\rho_i \in \mathbb{R}$ and
each $\gamma_j \in \mathbb{C} \setminus \mathbb{R}$. There are now
univariate polynomials $A_1, \ldots, A_{\ell}$ and $C_1, \ldots,
C_m$ such that, for all $n \geq 0$,
\[
u_n = \sum_{i=1}^{\ell} A_i(n) \rho_i^n + 
      \sum_{j=1}^m \left(C_j(n) \gamma_j^n + 
                  \overline{C_j}(n) \overline{\gamma_j}^n\right) \,.
\]

This expression is referred to as the \defemph{exponential polynomial}
solution of $\uu$. For integer LRS, the polynomials $A_i$ have real
algebraic coefficients and the polynomials $C_j$ have complex
algebraic coefficients. The degree of each of these polynomials is at
most one less than the multiplicity of the corresponding
characteristic root; thus in particular, these polynomials are
identically constant when $\uu$ has no repeated characteristic
roots. For fixed $k$, all coefficients appearing in these polynomials
can be computed in time polynomial in $||\uu||$, since they can be
obtained by solving a system of linear equations involving the first
$k$ terms of $\uu$. See Section~\ref{tools} for further details on
algebraic-number manipulations.

%Conversely, suppose that an infinite sequence $\uu = 
%\langle u_n \rangle_{n=0}^{\infty}$ is defined by an exponential
%polynomial as per Equation~(\ref{exp-poly}), where each $A_i \in
%\mathbb{R}[x]$ and each $C_j \in \mathbb{C}[x]$. Then $\uu$ is an 

An LRS is said to be \defemph{non-degenerate} if it does not have two
distinct characteristic roots whose quotient is a root of unity.  As
pointed out in~\cite{BOOK}, the study of arbitrary LRS can effectively
be reduced to that of non-degenerate LRS, by partitioning the original
LRS into finitely many subsequences, each of which is
non-degenerate. In general, such a reduction will require exponential
time. However, when restricting ourselves to LRS of bounded order (in
our case, of order at most $5$), the reduction can be carried out in
polynomial time. In particular, any LRS of order $5$ or less can be
partitioned in polynomial time into at most $2520$ non-degenerate LRS
of the same order or less.\footnote{We obtained this value using a
  bespoke enumeration procedure for order~$5$. A bound of $e^{2
    \sqrt{6\cdot 5 \log 5}} \leq 1,085,134$ can be obtained from
  Corollary~3.3 of~\cite{YLN95}.} In the rest of this paper, we shall
therefore assume that all LRS we are given are non-degenerate.

Any LRS $\uu$ of order $k$ can alternately be given in matrix form, in
the sense that there is a square matrix $M$ of dimension $k \times k$,
together with $k$-dimensional column vectors $\vec{v}$ and $\vec{w}$,
such that, for all $n \geq 0$, $u_n = \vec{v}^T M^n \vec{w}$. It
suffices to take $M$ to be the transpose of the companion matrix of
the characteristic polynomial of $\uu$, let $\vec{v}$ be the vector
$(u_{k-1},\ldots,u_0)$ of initial terms of $\uu$ in reverse order, and
take $\vec{w}$ to be the vector whose first $k-1$ entries are $0$ and
whose $k$th entry is $1$. It is worth noting that the characteristic
roots of $\uu$ correspond precisely to the eigenvalues of $M$. This
translation is instrumental in Section~\ref{decidability} to place the
Positivity Problem for LRS of order at most $5$ within the Counting
Hierarchy.

Conversely, given any square matrix $M$ of dimension $k \times k$, and
any $k$-dimensional vectors $\vec{v}$ and $\vec{w}$, let $u_n =
\vec{v}^T M^n \vec{w}$. Then $\langle \vec{v}^T M^n \vec{w}
\rangle_{n=k}^{\infty}$ is an LRS of order at most $k$ whose
characteristic polynomial is the same as that of $M$, as can be seen
by applying the Cayley-Hamilton Theorem.\footnote{In fact, if none of
  the eigenvalues of $M$ are zero, it is easy to see that the full
  sequence $\langle \vec{v}^T M^n \vec{w} \rangle_{n=0}^{\infty}$ is
  an LRS (of order at most $k$).}

Let $\langle u_n \rangle_{n=0}^{\infty}$ and $\langle v_n
\rangle_{n=0}^{\infty}$ be LRS of order $k$ and $l$
respectively. Their pointwise product $\langle u_n v_n
\rangle_{n=0}^{\infty}$ and sum $\langle u_n + v_n
\rangle_{n=0}^{\infty}$ are also LRS of order at most $kl$ and $k+l$
respectively. In the special case of pointwise squaring, the order of
the LRS $\langle u_n^2\rangle_{n=0}^{\infty}$ is at most $k(k+1)/2$.

We can use the above to reduce (the complement of) the Skolem Problem
to Positivity: given an integer LRS $\uu = \langle u_n
\rangle_{n=0}^{\infty}$, we see that $u_n \neq 0$ iff $u_n^2 -1 \geq
0$. Since this reduction is polynomial in $||\uu||$, the NP-hardness
for the Skolem Problem presented in~\cite{BP02} immediately translates
as coNP-hardness for Positivity, as pointed out in~\cite{BDJB10}. In
fact, since the LRS used in~\cite{BP02} are all periodic, we also
obtain a coNP-hardness for Ultimate Positivity. At the time of
writing, no other complexity bounds for these problems are known.

\section{Mathematical Tools}
\label{tools}

In this section we introduce the key technical tools used in this
paper.

For $p \in \mathbb{Z}[x_1, \ldots, x_m]$ a polynomial with integer
coefficients, let us denote by $||p||$ the bit length of its
representation as a list of coefficients encoded in binary. Note that
the degree of $p$ is at most $||p||$, and the height of $p$---i.e.,
the maximum of the absolute values of its coefficients---is at
most $2^{||p||}$.

We begin by summarising some basic facts about algebraic numbers and
their (efficient) manipulation. The main references
include~\cite{Coh93,BPR06,Ren92}.

A complex number $\alpha$ is \defemph{algebraic} if it is a root of a
single-variable polynomial with integer coefficients. The
\defemph{defining polynomial} of $\alpha$, denoted $p_{\alpha}$, is
the unique polynomial of least degree, and whose coefficients do not
have common factors, which vanishes at $\alpha$. The \defemph{degree}
and \defemph{height} of $\alpha$ are respectively those of
$p_{\alpha}$.

A standard representation\footnote{Note that this representation is
  not unique.} for algebraic numbers is to encode $\alpha$ as a
tuple comprising its defining polynomial together with rational
approximations of its real and imaginary parts of sufficient precision
to distinguish $\alpha$ from the other roots of $p_{\alpha}$. More
precisely, $\alpha$ can be represented by $(p_{\alpha},a, b,r) \in
\mathbb{Z}[x] \times \mathbb{Q}^3$ provided that $\alpha$ is the
unique root of $p_{\alpha}$ inside the circle in $\mathbb{C}$ of
radius $r$ centred at $a + bi $. A separation bound due to
Mignotte~\cite{Mig82} asserts that for roots $\alpha \neq \beta$ of a
polynomial $p \in \mathbb{Z}[x]$, we have
\begin{equation}
\label{root-sep-bound}
|\alpha - \beta| > \frac{\sqrt6}{d^{(d+1)/2} H^{d-1}} \, ,
\end{equation} 
where $d$ and $H$ are respectively the degree and height of $p$. Thus
if $r$ is required to be less than a quarter of the root-separation
bound, the representation is well-defined and allows for equality
checking. Given a polynomial $p \in \mathbb{Z}[x]$, it is well-known
how to compute standard representations of each of its roots in time
polynomial in $||p||$~\cite{Pan97,Coh93,BPR06}. Thus given $\alpha$ an
algebraic number for which we have (or wish to compute) a standard
representation, we write $||\alpha||$ to denote the bit length of this
representation. From now on, when referring to computations on
algebraic numbers, we always implicitly refer to their standard
representations. 

Note that Equation~(\ref{root-sep-bound}) can be used more generally
to separate arbitrary algebraic numbers: indeed, two algebraic numbers
$\alpha$ and $\beta$ are always roots of the polynomial
$p_{\alpha}p_{\beta}$ of degree at most the sum of the degrees of
$\alpha$ and $\beta$, and of height at most the product of the heights
of $\alpha$ and $\beta$.

Given algebraic numbers $\alpha$ and $\beta$, one can compute
$\alpha+\beta$, $\alpha \beta$, $1/\alpha$ (for non-zero $\alpha$),
$\overline{\alpha}$, and $|\alpha|$, all of which are algebraic, in
time polynomial in $||\alpha|| + ||\beta||$. Likewise, it is
straightforward to check whether $\alpha = \beta$. Moreover, if
$\alpha \in \mathbb{R}$, deciding whether $\alpha > 0$ can be done in
time polynomial in $||\alpha||$. Efficient algorithms for all these
tasks can be found in~\cite{Coh93,BPR06}.

Remarkably, integer multiplicative relationships among a fixed number
of algebraic numbers can be elicited systematically in polynomial
time:

\begin{theorem}
\label{Ge}
Let $m$ be fixed, and let $\lambda_1, \ldots, \lambda_m$ be complex
algebraic numbers of modulus $1$. Consider the free abelian group $L$
under addition given by
\[
L =  \{(v_1, \ldots, v_m) \in \mathbb{Z}^m : 
\lambda_1^{v_1} \ldots \lambda_m^{v_m} =
1\} \, .
\]
$L$ has a basis $\{\vec{\ell_1}, \ldots, \vec{\ell_p}\} \subseteq
\mathbb{Z}^m$ (with $p \leq m$), where the entries of each of the
$\vec{\ell_j}$ are all polynomially bounded in $||\lambda_1|| + \ldots
+ ||\lambda_m||$. Moreover, such a basis can be computed in time
polynomial in $||\lambda_1|| + \ldots +
||\lambda_m||$.
\end{theorem}

Note in the above that the bound is on the \emph{magnitude} of the
vectors $\vec{\ell_j}$ (rather than the bit length of their representation),
which follows from a deep result of Masser~\cite{Mas88}. For a proof
of Theorem~\ref{Ge}, see also~\cite{Ge93,CLZ00}.

We now turn to the first-order theory of the reals. Let $\vec{x} =
x_1, \ldots, x_m$ be a list of $m$ real-valued variables, and let
$\sigma(\vec{x})$ be a Boolean combination of
atomic predicates of the form $g(\vec{x}) \sim 0$, where each
$g(\vec{x}) \in \mathbb{Z}[\vec{x}]$ is a polynomial
with integer coefficients over these variables, and $\sim$ is either
$>$ or $=$.
%$g_1, \ldots, g_{\ell} \in \mathbb{Z}[\vec{x}]$ be polynomials with
%integer coefficients over these variables. Let $\sigma(\vec{x})$ be a
%Boolean combination of atomic predicates of the form $g_i(\vec{x})
%\sim_i 0$, where each $\sim_i$ is either $>$ or $=$.  
A \defemph{sentence of the first-order theory of the reals} is of the
form
\begin{equation}
\label{for-formula}
Q_1 x_1 \ldots Q_m x_m \, \sigma (\vec{x}) \, ,
\end{equation}
where each $Q_i$ is one of the quantifiers $\exists$ or $\forall$.
Let us denote the above formula by $\tau$, and write
$||\tau||$ to denote the bit length of its syntactic
representation.

Tarski famously showed that the first-order theory of the reals is
decidable~\cite{Tar51}. His procedure, however, has non-elementary
complexity. Many substantial improvements followed over the years,
starting with Collins's technique of cylindrical algebraic
decomposition~\cite{Col75}, and culminating with the fine-grained
analysis of Renegar~\cite{Ren92}. In this paper, we focus exclusively
on the situation in which the number of variables is uniformly
bounded.

\begin{theorem}[Renegar]
\label{renegar}
Let $M \in \mathbb{N}$ be fixed. Let $\tau$ be of the
form~(\ref{for-formula}) above. Assume that the number of variables in
$\tau$ is bounded by $M$ (i.e., $m \leq M$). Then the truth value of
$\tau$ can be determined in time polynomial in $||\tau||$.
\end{theorem}

Theorem~\ref{renegar} follows immediately from~\cite[Thm.~1.1]{Ren92}.

Our next result is a special case of Kronecker's famous theorem on
simultaneous Diophantine approximation, a statement and proof of
which can be found in~\cite[Chap.~7, Sec.~1.3, Prop.~7]{Bou66}.

For $x \in \mathbb{R}$, write $[x]_{2 \pi}$ to denote the distance
from $x$ to the closest integer multiple of $2 \pi$: $[x]_{2 \pi} =
\min \{ |x - 2\pi j| : j \in \mathbb{Z} \}$.

\begin{theorem}[Kronecker]
\label{kronecker}
Let $t_1, \ldots, t_m, x_1, \ldots, x_m \in [0, 2\pi)$. The following
  are equivalent:
\begin{enumerate}
\item 
For any $\varepsilon > 0$, there exists $n \in \mathbb{Z}$ such
  that, for $1 \leq j \leq m$, we have $[n t_j - x_j]_{2\pi} 
\leq \varepsilon$.

\item For every tuple $(v_1, \ldots, v_m)$ of integers such that $[v_1
  t_1 + \ldots + v_m t_m]_{2\pi} = 0$, we have
$[v_1 x_1 + \ldots + v_m x_m]_{2\pi} = 0$.
\end{enumerate}
\end{theorem}

We can strengthen Theorem~\ref{kronecker} by requiring that $n \in
\mathbb{N}$ in the first assertion. Indeed, suppose that in a given
instance, we find that $n < 0$. A straightforward pigeonhole argument
shows that there exist arbitrarily large positive integers $g$ such
that $[g t_j]_{2\pi} \leq \varepsilon$ for $1 \leq j \leq m$. It
follows that $[(g+n) t_j - x_j]_{2 \pi} \leq 2 \varepsilon$, which
establishes the claim for sufficiently large $g$ (noting that
$\varepsilon$ is arbitrary).

Let $\lambda_1, \ldots, \lambda_m$ be complex algebraic numbers of
modulus $1$. For each $j \in \{1, \ldots, m\}$, write $\lambda_j =
e^{i\theta_j}$ for some $\theta_j \in [0, 2\pi)$.  Let 
\begin{align*}
L = & \ \{(v_1,
  \ldots, v_m) \in \mathbb{Z}^m : \lambda_1^{v_1} \ldots
  \lambda_m^{v_m} = 1\} \\
  = & \  \{(v_1, \ldots, v_m) \in
  \mathbb{Z}^m : [v_1 \theta_1 + \ldots + v_m \theta_m]_{2\pi} =
  0\} \, .
\end{align*}
Recall from Theorem~\ref{Ge} that $L$ is a free abelian group under
addition with basis $\{\vec{\ell_1}, \ldots, \vec{\ell_p}\} \subseteq
\mathbb{Z}^m$, where $p \leq m$.

For each $j \in \{1, \ldots, p\}$, let $\vec{\ell_j} = (\ell_{j,1},
\ldots, \ell_{j,m})$.  Write
\begin{align*}
R = & \ \{ \vec{x} = (x_1, \ldots, x_m) \in [0,2\pi)^m : \\
& \ \ \, [\vec{\ell_j} \cdot \vec{x}]_{2\pi} = 0 \mbox{ for } 1 \leq j \leq
p\} \, .
\end{align*}
By Theorem~\ref{kronecker}, for an arbitrary tuple $(x_1, \ldots,
x_m) \in [0,2\pi)^m$, it is the case that, for all $\varepsilon > 0$,
  there exists $n \in \mathbb{N}$ such that, for $j \in \{1, \ldots,
  m\}$, $[n \theta_j - x_j]_{2\pi} \leq \varepsilon$ iff
  $(x_1, \ldots, x_m) \in R$. 

Now observe that $(x_1, \ldots, x_m) \in R$ iff $(e^{i
  x_1}, \ldots, e^{i x_m}) \in T$, where
\begin{align*}
T = & \ \{(z_1, \ldots, z_m) \in \mathbb{C}^m
: |z_1| = \ldots = |z_m| = 1 
\mbox{ and,} \\
&\ \ \, \mbox{for each $j \in \{1, \ldots, p\}$, }
z_1^{\ell_{j,1}} \ldots z_m^{\ell_{j,m}} = 1 \} \, .
\end{align*}

Since $e^{i n \theta_j} = \lambda_j^n$,
we immediately have the following:

\begin{corollary}
\label{density}
Let $\lambda_1, \ldots, \lambda_m$ and $T$ be as above. Then
$\{(\lambda_1^n, \ldots, \lambda_m^n) : n \in \mathbb{N}\}$ is
a dense subset of $T$.
\end{corollary}

Finally, we give a version of Baker's deep theorem on linear forms in
logarithms. The particular statement we have chosen is a sharp formulation
due to Baker and W\"ustholz~\cite{BW93}.

In what follows, $\log$ refers to the principal value of the complex
logarithm function given by $\log z = \log |z| + i \arg z$, where
$-\pi < \arg z \leq \pi$.

\begin{theorem}[Baker and W\"ustholz]
\label{Baker}
Let $\alpha_1, \ldots, \alpha_m \in \mathbb{C}$ be algebraic numbers
different from $0$ or $1$, and let $b_1,\ldots,b_m \in \mathbb{Z}$ be
integers. Write
\[\Lambda = b_1 \log \alpha_1 + \ldots + b_m \log \alpha_m \, . \]
Let $A_1, \ldots, A_m, B \geq e$ be real numbers such that, for each
$j \in \{1, \ldots, m\}$, $A_j$ is an upper bound for the height of
$\alpha_j$, and $B$ is an upper bound for $|b_j|$. Let $d$ be the
degree of the extension field $\mathbb{Q}(\alpha_1, \ldots, \alpha_m)$
over $\mathbb{Q}$.

If $\Lambda \neq 0$, then 
\[
\log |\Lambda| > -(16md)^{2(m+2)}\log A_1
\ldots \log A_m \log B \, .
\]  
\end{theorem}

Finally, we record the following fact, whose straightforward proof is
left to the reader.

\begin{proposition}
\label{exp-bound}
Let $a \geq 2$ and $\varepsilon \in (0,1)$ be real numbers. Let $B \in
\mathbb{Z}[x]$ have degree at most $a^{D_1}$ and height at most
$2^{a^{D_2}}$, and assume that $1/\varepsilon \leq 2^{a^{D_3}}$, for
    some $D_1,D_2,D_3 \in \mathbb{N}$. Then there is $D_4 \in
    \mathbb{N}$ depending only on $D_1, D_2, D_3$ such that, for all
    $n \geq 2^{a^{D_4}}$, $\displaystyle{\frac{1}{B(n)} >
 (1-\varepsilon)^n}$.
\end{proposition}

\section{Decidability and Complexity}
\label{decidability}

Let $\uu = \langle u_n \rangle_{n=0}^{\infty}$ be an integer LRS of
order $k$. 
As discussed in the Introduction, we assume that $u$ is
presented as a $2k$-tuple of integers $(a_1, \ldots, a_k, u_0, \ldots,
u_{k-1}) \in \mathbb{Z}^{2k}$, such that for all $n \geq 0$,
\begin{equation}
\label{recurrence}
u_{n+k} = a_1 u_{n+k-1} + \ldots + a_k u_n \, .
\end{equation}
%and write $||\uu||$ to denote the bit length of the tuple $(a_1,\ldots,
%a_k, u_0, \ldots, u_{k-1})$ encoded in binary.

The \defemph{Positivity Problem} asks, given such an LRS $\uu$,
whether for all $n \geq 0$, it is the case that $u_n \geq 0$. When
this holds, we say that $\uu$ is \defemph{positive}.

The \defemph{Ultimate Positivity Problem} asks, given such an LRS
$\uu$, whether there exists $N \geq 0$ such that, for all $n \geq N$,
it is the case that $u_n \geq 0$. When this holds, we say that $\uu$
is \defemph{ultimately positive}.

In this section, we establish the following main results:

\begin{theorem}
\label{theorem-pos}
The Positivity Problem for integer LRS of order $5$ or less is
decidable in $\mathrm{coNP}^{\mathrm{PP}^{\mathrm{PP}^{\mathrm{PP}}}}$.
\end{theorem}

\begin{theorem}
\label{theorem-upos}
The Ultimate Positivity Problem for integer LRS of order $5$ or
  less is decidable in polynomial time.
\end{theorem}

Note that the above results immediately carry over to rational
LRS\@. To see this, consider a rational LRS $\uu$ obeying the
recurrence relation~(\ref{recurrence}). Let $\ell$ be the least common
multiple of the denominators of the rational numbers $a_1, \ldots,
a_k, u_0, \ldots, u_{k-1}$, and define an integer sequence $\vv =
\langle v_n \rangle_{n=0}^{\infty}$ by setting $v_n = \ell^{n+1} u_n$
for all $n \geq 0$. It is easily seen that $\vv$ is an integer LRS of
the same order as $\uu$, and that for all $n$, $v_n \geq 0$ iff $u_n
\geq 0$. 

\

\noindent
\textbf{Positivity---High-Level Synopsis.}  At a high level, the
algorithm upon which Theorem~\ref{theorem-pos} rests proceeds as
follow. Given an LRS $\uu$, we first decide whether or not $\uu$ is
ultimately positive by studying its exponential polynomial
solution---further details on this task are provided shortly. As we
prove in this paper, whenever $\uu$ is an ultimately positive LRS of
order~$5$ or less, there is an effective bound $N$ of at most
exponential magnitude such that all terms of $\uu$ beyond $N$ are
positive. Next, observe that $\uu$ cannot be positive unless it is
ultimately positive. Now in order to assert that an ultimately
positive LRS $\uu$ is \emph{not} positive, we use a
\emph{guess-and-check} procedure: find $n \leq N$ such that $u_n <
0$. By writing $u_n = \vec{v}^T M^n \vec{w}$, for some square integer
matrix $M$ and vectors $\vec{v}$ and $\vec{w}$
(cf.~Section~\ref{sec-LRS}), we can decide whether $u_n < 0$ in
$\mathrm{PosSLP}$\footnote{Recall that $\mathrm{PosSLP}$ is the
  problem of determining whether an arithmetic circuit, with addition,
  multiplication, and subtraction gates, evaluates to a positive
  integer.} via iterative squaring, which yields an
$\mathrm{NP}^{\mathrm{PosSLP}}$ procedure for non-Positivity. Thanks
to the work of Allender \emph{et al.}~\cite{ABK09}, which asserts that
$\mathrm{PosSLP} \subseteq
\mathrm{P}^{\mathrm{PP}^{\mathrm{PP}^{\mathrm{PP}}}}$, we obtain the
required $\mathrm{coNP}^{\mathrm{PP}^{\mathrm{PP}^{\mathrm{PP}}}}$
algorithm for deciding Positivity.

\

The following is an old result concerning LRS; proofs can be found
in~\cite[Thm.~7.1.1]{GL91} and \cite[Thm.~2]{BG07}. It also follows
easily and directly from either Pringsheim's theorem or
from~\cite[Lem.~4]{Bra06}. It plays an important role in our approach
by enabling us to significantly cut down on the number of subcases
that must be considered, avoiding the sort of quagmire alluded to
in~\cite{NS90}.

\begin{proposition}
\label{prune}
Let $\langle u_n \rangle_{n=0}^{\infty}$ be an LRS with no real
positive dominant characteristic root. Then there are infinitely many
$n$ such that $u_n < 0$ and infinitely many $n$ such that $u_n > 0$.
\end{proposition}

By Proposition~\ref{prune}, it suffices to restrict our attention to
LRS whose dominant characteristic roots include one real positive
value. Given an integer LRS $\uu$, note that determining whether the
latter holds is easily done in time polynomial in $||\uu||$.

Thus let $\uu$ be a non-degenerate integer LRS of order $k$ having a
(possibly repeated) real positive dominant characteristic root $\rho >
0$. Note that $\uu$ cannot have a real negative dominant
characteristic root (which would be $-\rho$), since otherwise the
quotient $-\rho/\rho = -1$ would be a root of unity, contradicting
non-degeneracy.  Let us therefore write the characteristic roots as
$\{\rho, \gamma_1, \overline{\gamma_1}, \ldots, \gamma_m,
\overline{\gamma_m}\} \cup \{\gamma_{m+1}, \gamma_{m+2}, \ldots,
\gamma_{\ell}\}$, where we assume that the roots in the first set all
have common modulus $\rho$, whereas the roots in the second set all
have modulus strictly smaller than $\rho$. Note that for LRS of order
at most $5$, $m$ can be at most $2$.

Let $\lambda_i = \gamma_i/\rho$ for $1 \leq i \leq \ell$. We can
then write
\begin{equation}
\label{LRSrep1}
\frac{u_n}{\rho^n} = A(n) + 
          \sum_{i=1}^m \left(C_i(n) \lambda_i^n + 
          \overline{C_i}(n) \overline{\lambda_i}^n\right)+ r(n) \, ,
\end{equation}
for a suitable real polynomial $A$ and complex polynomials $C_1,
\ldots, C_m$, where $r(n)$ is a term tending to zero
exponentially fast. 

Note that none of $\lambda_1, \ldots, \lambda_m$, all of which have
modulus $1$, can be a root of unity, as each $\lambda_i$ is a quotient
of characteristic roots and $\uu$ is assumed to be
non-degenerate. 

For $i \in \{1, \ldots, \ell\}$, observe also that as each $\lambda_i$
is a quotient of two roots of the same polynomial of degree $k$, it
has degree at most $k(k-1)$. In fact, it is easily seen that
$||\lambda_i|| = ||\uu||^{\mathcal{O}(1)}$.

As noted in Section~\ref{sec-LRS}, the degree of polynomials $A$
and $C_i$ in the exponential polynomial solution is at most one less
than the multiplicity of the corresponding characteristic roots, and
is therefore bounded above by $k-1$. Recall also that all coefficients
appearing in these polynomials are algebraic and, for fixed $k$, can
be computed and manipulated in time polynomial in $||\uu||$. It easily
follows that $||A|| = ||\uu||^{\mathcal{O}(1)}$ and $||C_i|| =
||\uu||^{\mathcal{O}(1)}$.

Finally, we place bounds on the rate of convergence of $r(n)$.
We have 
\[
r(n) = C_{m+1}(n) \lambda_{m+1}^n + \ldots +
C_{\ell}(n)\lambda_{\ell}^n \, .
\]
For fixed $k$, combining our estimates
on the height and degree of each $\lambda_i$ together with the
root-separation bound given by Equation~(\ref{root-sep-bound}), we get
$\left|\frac{1}{1-\lambda_i}\right| =
  2^{||\uu||^{\mathcal{O}(1)}}$, for $m+1 \leq i \leq \ell$. Thanks
also to the bounds on the height and degree of the polynomials $C_i$,
it follows that we can find $\varepsilon \in (0,1)$ and $N \in
\mathbb{N}$ such that:
\begin{align}
\label{eq1}
& 1/\varepsilon = 2^{||\uu||^{\mathcal{O}(1)}} \\
& N = 2^{||\uu||^{\mathcal{O}(1)}}  \\
& \mbox{For all } n > N,\ |r(n)| < (1-\varepsilon)^n \, .
\label{eq3}
\end{align}
In addition, we can compute such $\varepsilon$ and $N$ in time
polynomial in $||\uu||$. Naturally, given $k$, we can also assume that
we have calculated explicitly once and for all the constants implicit
in the various instances of the $\mathcal{O}(1)$ notation.

We now seek to answer Positivity and Ultimate Positivity for the LRS
$\uu = \langle u_n \rangle_{n=0}^{\infty}$ by studying the same for
$\langle u_n/\rho^n \rangle_{n=0}^{\infty}$.

In what follows, we assume that $\uu$ is as given above; in
particular, $\uu$ is a non-degenerate integer LRS having a (possibly
repeated) real positive dominant characteristic root $\rho > 0$.

\

\noindent
\textbf{Ultimate Positivity---High-Level Synopsis.}  
Before launching into technical details, let us provide a high-level
overview of our proof strategy for deciding Ultimate Positivity.
Consider first the special case of Equation~(\ref{LRSrep1}) in which
the polynomials $A(n)$ and $C_i(n), \overline{C_i}(n)$ are all
identically constant. Let us rewrite this equation as
\begin{equation}
\label{eq-explanation}
\frac{u_n}{\rho^n} = A + h(\lambda_1^n, \ldots, \lambda_m^n) 
                              + r(n) \, ,
\end{equation}
where $h:\mathbb{C}^m \rightarrow \mathbb{R}$ is a continuous
function. In general, there will be integer multiplicative
relationships among the $\lambda_1, \ldots, \lambda_m$, for which we
can compute a basis $B$ thanks to Theorem~\ref{Ge}. These
multiplicative relationships define a torus $T \subseteq \mathbb{C}^m$
on which the joint iterates $(\lambda_1^n, \ldots, \lambda_m^n)$
are dense, as per Kronecker's theorem
(in the form of Corollary~\ref{density}).

If $r(n)$ is identically $0$, then both Positivity and Ultimate
Positivity can be decided by determining the sign of the expression $A
+ \min h\mbox{$\restriction$}_T$ (where $h\mbox{$\restriction$}_T$
denotes the function $h$ restricted to the torus $T$). For fixed order
$k$, computing this sign can be carried out in polynomial time via the
first-order theory of the reals, thanks to Theorem~\ref{renegar}.

If $r(n)$ is not identically $0$, then for LRS of order at most $5$,
we have that $m$ is either $0$ or $1$, where the latter is the
interesting case. The torus $T$ is now simply the unit circle in the
complex plane, and Equations~(\ref{LRSrep1}) and (\ref{eq-explanation})
can be rewritten as
\[
\frac{u_n}{\rho^n} = A + 2|c_1|\cos(n \theta_1 + \varphi_1) + r(n) \, ,
\]
where $C_1(n) = c_1 = |c_1|e^{i\varphi_1}$ and $\theta_1 = \arg
\lambda_1$. The critical case now arises when $A-2|c_1| = 0$, which we
can determine in polynomial time. Noting that the cosine function is
minimised when its argument is an odd integer multiple of $\pi$, we
can use Baker's theorem to bound the expression $n \theta_1 + \varphi_1$
away from odd integer multiples of $\pi$ by an inverse polynomial in
$n$. Using a Taylor approximation, we then argue that $\cos(n \theta_1
+ \varphi_1)$ is itself eventually bounded away from $-1$ by a
(different) inverse polynomial in $n$, and since $r(n)$ decays to zero
exponentially fast, we are able to conclude that $u_n/\rho^n$ is
ultimately positive, and can compute a bound $N$ after which all terms
$u_n$ (for $n > N$) are positive.

Returning to Equation~(\ref{LRSrep1}), note that if the
$C_i(n),\overline{C_i}(n)$ are all identically constant but $A(n)$ is
not, then the latter will eventually dominate and enable us to settle
the ultimate positivity question; likewise, if $A(n)$ is identically
constant but some $C_i(n),\overline{C_i}(n)$ are not, the latter
eventually dominate and the situation can be dealt with straightforwardly.

This analysis allows us to handle LRS of order up to $5$. At order
$6$, however, we encounter a critical situation in which $A(n)$ and
$C_1(n),\overline{C_1}(n)$ are all linear polynomials, which then
leads to the hardness results described in
Section~\ref{hardness}. 

\

We now proceed with the proofs of Theorems~\ref{theorem-pos} and
\ref{theorem-upos}, split into cases according to the number of
distinct (albeit possibly repeated) dominant characteristic roots of
$\uu$. Since there is one real positive dominant root, no real
negative dominant root, and since non-real roots always arise in
pairs, the number of dominant roots must be odd. In any event, the
total number of characteristic roots is bounded by the order of $\uu$,
which we assume to be at most $5$.

\subsection{One Dominant Root.}
In case of a single dominant root $\rho \in \mathbb{R}$, from
Equation~(\ref{LRSrep1}) we have that ${u_n/\rho^n} = A(n) + r(n)$. If
$A(n)$ is identically $0$, we simply turn our attention towards
$r(n)$, which is an LRS whose exponential polynomial solution has one
fewer term. Otherwise, it is clear that $\uu$ is ultimately positive
iff either $A(n)$ is identically equal to some constant $a > 0$, or
$\lim_{n \rightarrow \infty}A(n) = \infty$, all of which can be
decided straightforwardly in time polynomial in $||\uu||$.

Turning to positivity, assume therefore that $A(n)$ is either a
strictly positive constant or tends to $\infty$. Recall from our
earlier discussion on the rate of convergence of $r(n)$ that we can
compute in polynomial time numbers $\varepsilon \in (0,1)$, with
$1/\varepsilon = 2^{||\uu||^{\mathcal{O}(1)}}$, and $N =
2^{||\uu||^{\mathcal{O}(1)}}$, such that $|r(n)| < (1-\varepsilon)^n$
for all $n > N$. We can similarly compute a bound $N' =
2^{||\uu||^{\mathcal{O}(1)}}$ such that $A(n) \geq (1-\varepsilon)^n$
for all $n > N'$. Let $N'' = \max\{N,N'\}$. Then $\uu$ will fail to be
positive iff there is some $n \leq N''$ such that $u_n < 0$. Since
$N''$ is at most exponential in $||\uu||$, we can decide positivity of
$\uu$ in $\mathrm{coNP}^{\mathrm{PP}^{\mathrm{PP}^{\mathrm{PP}}}}$ via
a $\mathrm{PosSLP}$ oracle as outlined earlier.

\subsection{Three Dominant Roots.}
Next, we consider the case in which $\uu$ has exactly three dominant
characteristic roots $\{\rho, \gamma_1, \overline{\gamma_1}\}$. Two
subcases arise: (i)~either the complex roots $\gamma_1$ and
$\overline{\gamma_1}$ are simple, or (ii)~$\gamma_1$ and
$\overline{\gamma_1}$ are repeated.

(i)~In the first subcase, the multiplicity of $\rho$ may range from
$1$ to $3$. If $\rho$ has multiplicity $3$ then there can be no other
characteristic roots, and 
\[
u_n/\rho^n = an^2 + bn + d+ c_1 \lambda_1^n
+ \overline{c_1} \overline{\lambda_1}^n \, ,
\] 
where $a,b,d$ are real algebraic constants, and $c_1$ is a complex
algebraic constant which we assume is non-zero (otherwise the
situation is trivial).

If $a < 0$, then clearly $\uu$ is neither positive nor ultimately
positive. If $a > 0$ then $\uu$ is ultimately positive and, similarly
to the case of a single dominant root, we can use our earlier
estimates on the height and degree of $a$, $b$, $d$, and $c_1$,
together with the root-separation bound given by
Equation~(\ref{root-sep-bound}), to conclude that there is $N =
2^{||\uu||^{\mathcal{O}(1)}}$ such that, for all $n > N$, we have $u_n
\geq 0$. The positivity of $\uu$ can then be decided in 
$\mathrm{coNP}^{\mathrm{PP}^{\mathrm{PP}^{\mathrm{PP}}}}$.

Next, if $a = 0$ then there is potentially an exponentially decaying
term in the exponential polynomial solution of $u_n/\rho^n$;
this also covers the case in which the multiplicity of $\rho$ is $1$
or $2$: 
\[
u_n/\rho^n = bn + d+ c_1 \lambda_1^n + \overline{c_1}
\overline{\lambda_1}^n + r(n) \, .
\]

Here, similarly to the previous case, if $b < 0$ then $\uu$ is neither
positive nor ultimately positive, whereas if $b > 0$ then $\uu$ is
ultimately positive and, as before, we obtain an exponential upper
bound on the index $n$ of possible violations of positivity, as
required.

Finally, suppose that $a=0$ and $b=0$. We may assume that $c_1 \neq
0$, otherwise we are left with the term $r(n)$ and can simply recast
our analysis appropriately at lower order. Let $\theta_1 = \arg
\lambda_1$ and $\varphi_1 = \arg c_1$.  We have
\[ \frac{u_n}{\rho^n} = d + 2 |c_1| \cos(n \theta_1 + \varphi_1) + r(n) \,
. \] 
Since $\lambda_1$ is not a root of unity, it is straightforward to see
that the set $\{\cos(n \theta_1 + \varphi_1) : n \geq 0\}$ is dense in
$[-1,1]$. It immediately follows that if $d < 2 |c_1|$ then $\uu$ is
neither positive nor ultimately positive, whereas if $d > 2 |c_1|$
then $\uu$ is ultimately positive with, as before, an exponential
bound on the index of possible violations of positivity.

It remains to tackle the case in which $d = 2 |c_1|$.  Since
$\lambda_1$ is not a root of unity, there is at most one value of $n$
such that $n \theta_1 + \varphi_1$ is an odd integer multiple of
$\pi$, corresponding to $\lambda_1^n = -|c_1|/c_1$. It then follows
from Theorem~\ref{Ge} that this value (if it exists) is at most $M =
||\uu||^{\mathcal{O}(1)}$.

By Equations~(\ref{eq1})--(\ref{eq3}), we can find $\varepsilon \in
(0,1)$ and $N = 2^{||\uu||^{\mathcal{O}(1)}}$ such that for all $n >
N$, we have $|r(n)| < (1-\varepsilon)^n$, and moreover $1/\varepsilon
= 2^{||\uu||^{\mathcal{O}(1)}}$.

Let $g(x) = \displaystyle{\frac{x^2}{2!} - \frac{x^4}{4!}}$. Using
a Taylor approximation, we have the following:
\begin{alignat}{2}
& \cos(x+\pi) \geq -1 + g(x) &  & \qquad \mbox{for $x \in (-\pi,\pi]$} \\
& g(x) \leq g(y) & & \qquad \mbox{for $|x| \leq |y| \leq 1$} \\
& 11/24 = g(1) \leq g(x) & & \qquad \mbox{for $1 \leq |x|
    \leq \pi$} \, .
\label{eqq3}
\end{alignat}

For $n \in \mathbb{N}$, write $\Lambda(n) = n \theta_1 + \varphi_1 -
(2j+1) \pi$, where $j \in \mathbb{Z}$ is the unique integer such that
$-\pi < \Lambda(n) \leq \pi$. We now have:
\begin{align*}
\smash{\frac{u_n}{\rho^n}} = & \ 2|c_1| + 2|c_1|\cos(n \theta_1 + \varphi_1) 
                       + r(n) \\
= & \ 2|c_1|(1 + \cos(\Lambda(n) + \pi)) + r(n) \\
  \geq & \ 2|c_1|g(\Lambda(n)) - (1-\varepsilon)^n \, ,
\end{align*}
where the inequality holds provided that $n > N$.

By Equation~(\ref{eqq3}), when $|\Lambda(n)| \geq 1$, we have
$\displaystyle{ \frac{u_n}{\rho^n} \geq \frac{11}{12}|c_1| -
  (1-\varepsilon)^n }$. It follows easily that $u_n/\rho^n \geq 0$
provided that $|\Lambda(n)| \geq 1$ and $n > N'$, for some $N' =
2^{||\uu||^{\mathcal{O}(1)}}$.

Recall that for $n > M$, $n\theta_1 + \varphi_1$ can never be an odd
integer multiple of $\pi$, i.e., $\Lambda(n) \neq 0$. We now claim
that there is an absolute constant $K \in \mathbb{N}$ such that, for
$n > M$, we have $|\Lambda(n)| > n^{-||\uu||^K}$.

To see this, write 
\[ \Lambda(n) = \frac{1}{i} 
\left( n \log \lambda_1 + \log \frac{c_1}{|c_1|} -(2j+1) \log (-1)
\right) \, .
\]
In the above, if $c_1 \in \mathbb{R}$ and $c_1 > 0$, then simply
remove the term $\displaystyle{\log \frac{c_1}{|c_1|} = 0}$ from the
expression for $\Lambda(n)$, which would yield an even better lower
bound than is obtained below. We may therefore assume without loss of
generality that $\lambda_1$ and $c_1/|c_1|$ are different from $0$ and
$1$.

Let $H \geq e$ be an upper bound for the heights of $\lambda_1$ and
$c_1/|c_1|$, and let $D$ be the largest of the degrees of $\lambda_1$
and $c_1/|c_1|$. Notice that the degree of
$\mathbb{Q}(\lambda_1,c_1/|c_1|)$ over $\mathbb{Q}$ is at most $D^2$,
and that $|j| \leq n$. We can thus invoke Theorem~\ref{Baker} to
conclude that
\begin{align*}
|\Lambda(n)| > & \ \exp 
\left( -(48 D^2)^{10}\log^2 H \log (2n+1) \right)\\ 
= & \
\frac{1}{(2n+1)^{(\log^2 H)(48 D^2)^{10}}}
\, ,
\end{align*}
for $n > M$. The claim now follows by noting that both $\log H$
and $D$ are bounded above by 
$||\lambda_1|| + ||(c_1/|c_1|)||$, and that the latter is in
$\mathcal{O}(||u||)$.

Thus when $|\Lambda(n)| < 1$ (and $n > M$), we have 
$g(\Lambda(n)) \geq g(n^{-||\uu||^K})$. We can therefore find a
polynomial $B \in \mathbb{Z}[x]$ such that
\[ 2|c_1|g(\Lambda(n)) \geq \frac{1}{B(n)} \, ,
\] 
requiring in addition that $B$ have degree $||\uu||^{\mathcal{O}(1)}$
and height $2^{||\uu||^{\mathcal{O}(1)}}$, where the latter is
achieved via bounds on the height of $|c_1|$. We can now invoke
Proposition~\ref{exp-bound} to conclude that there is $N'' =
2^{||\uu||^{\mathcal{O}(1)}}$ such that $N'' \geq M$ and, for all $n
> N''$, we have $\frac{1}{B(n)} > (1 - \varepsilon)^n$. Combining
our various inequalities, we see that $u_n/\rho^n \geq 0$
provided that $n > \max\{N,N',N''\}$, which establishes ultimate
positivity of $\uu$ and moreover once again provides an exponential
bound on the index of possible violations of positivity, as required.

This concludes Subcase~(i).

(ii)~Finally, we turn to the situation in which the complex dominant
roots $\gamma_1$ and $\overline{\gamma_1}$ are repeated. Using the
same notation as above, we have
\begin{align*}
\smash{\frac{u_n}{\rho^n}} = & \ a + (c_1n + c)\lambda_1^n + 
        (\overline{c_1}n + \overline{c})\overline{\lambda_1}^n \\ 
= & \ a + n(c_1 \lambda_1^n +
\overline{c_1} \overline{\lambda_1}^n) + c \lambda_1^n + \overline{c}
\overline{\lambda_1}^n \, .
\end{align*}

Note that, unless $c_1 = 0$, the term $c_1 \lambda_1^n +
\overline{c_1} \overline{\lambda_1}^n = 2 |c_1| \cos(n \theta_1 +
\varphi_1)$ is infinitely often negative and bounded away from zero,
which immediately entails that $\uu$ can be neither positive nor
ultimately positive. If $c_1 = 0$, on the other hand, we simply revert
to an instance considered under Subcase~(i).

\subsection{Five Dominant Roots.}
\label{5-dom-roots}
If an LRS of order $5$ has $5$ distinct dominant roots, then each root
is simple, and in Equation~(\ref{LRSrep1}) we have that $m=2$, $r(n)$ is
identically $0$, and the polynomials $A(n)$, $C_1(n)$, and $C_2(n)$ are
all identically constant (cf.~Section~\ref{sec-LRS}): 
\[
\frac{u_n}{\rho^n} = a + c_1 \lambda_1^n + 
\overline{c_1} \overline{\lambda_1}^n + 
c_2 \lambda_2^n + 
\overline{c_2} \overline{\lambda_2}^n \, ,
\]
for algebraic constants $a \in \mathbb{R}$ and
$c_1, c_2 \in \mathbb{C}$.

Let $L = \{(v_1, v_2) \in \mathbb{Z}^2 : \lambda_1^{v_1}
\lambda_2^{v_2} = 1\}$, and let $B$ be a basis for $L$. Note that $B$
can only have cardinality $0$ (when $L$ is trivial) or $1$, since it
is easily seen that the presence of two non-trivial independent
integer multiplicative relationships over $\lambda_1$ and $\lambda_2$
would entail that $\lambda_1$ and $\lambda_2$ are roots of unity,
contradicting the non-degeneracy of $\uu$. Recall from
Theorem~\ref{Ge} that the basis $B$ can be computed in polynomial
time, and moreover that elements of $B$ may be assumed to have
magnitude polynomial in $||\uu||$.

If $B = \emptyset$, let
\[
T = \{(z_1, z_2) \in \mathbb{C}^2 : |z_1| = |z_2| = 1\} \, ,
\]
and if $B = \{(\ell_1,\ell_2)\}$, write
\[
T = \{(z_1, z_2) \in \mathbb{C}^2 : |z_1| = |z_2| = 1 
\mbox{ and } z_1^{\ell_1}z_2^{\ell_2} = 1\} \, .
\]

Define $h : T \rightarrow \mathbb{R}$ by setting 
\[
h(z_1, z_2) = c_1 z_1 + \overline{c_1} \overline{z_1} + 
c_2 z_2 + \overline{c_2}
\overline{z_2} \, ,
\]
so that for all $n$, we have $u_n/\rho^n = a + h(\lambda_1^n,
\lambda_2^n)$. By Corollary~\ref{density}, the set $\{(\lambda_1^n,
\lambda_2^n) : n \in \mathbb{N}\}$ is a dense subset of $T$. Since $h$
is continuous, we immediately have that
\[
\inf \{u_n/\rho^n : n \in
\mathbb{N}\} = \min \{a + h(z_1, z_2) : (z_1,z_2) \in T\} \, .
\]
It follows that $\uu$ is ultimately positive iff $\uu$ is positive iff
$\min \{a + h(z_1,z_2) : (z_1,z_2) \in T\} \geq 0$ iff
\begin{equation}
\label{min-eqn}
 \forall (z_1, z_2) \in T,\, a + h(z_1, z_2) \geq 0 \, .
\end{equation}

We now show how to rewrite Assertion~(\ref{min-eqn}) as a sentence in
the first-order theory of the reals, i.e., involving only real-valued
variables and first-order quantifiers, Boolean connectives, and
integer constants together with the arithmetic operations of addition,
subtraction, multiplication, and division.\footnote{In
  Section~\ref{tools}, we did not include division as an allowable
  operation when we introduced the first-order theory of the reals;
  however instances of division can always be removed in linear time
  at the cost of introducing a linear number of existentially
  quantified fresh variables.} The idea is to separately represent the
real and imaginary parts of each complex quantity appearing in
Assertion~(\ref{min-eqn}), and combine them using real arithmetic so
as to mimic the effect of complex arithmetic operations.

To this end, we use pairs of real variables $x_1,y_1$ and $x_2,y_2$ to
represent $z_1$ and $z_2$ respectively: intuitively, $z_1 = x_1 + i
y_1$ and $z_2 = x_2 + i y_2$. Since the real constant $a$ is
algebraic, there is a formula $\sigma_a(x)$ which is true over the
reals precisely for $x = a$. Likewise, the real and imaginary parts
$\mathrm{Re}(c_1)$, $\mathrm{Im}(c_1)$, $\mathrm{Re}(c_2)$, and
$\mathrm{Im}(c_2)$ of the complex algebraic constants $c_1$ and $c_2$
are themselves real algebraic, and can be represented as
single-variable formulas in the first-order theory of the reals. All
such formulas can readily be shown to have size polynomial in $||\uu||$.

The terms $z_1^{\ell_1}$ and $z_2^{\ell_2}$ (if present) are simply
expanded: for example, if $\ell_1$ is positive, then $z_1^{\ell_1} =
(x_1 + i y_1)^{\ell_1} = A_1(x_1) + i B_1(y_1)$, where $A_1$ and $B_1$
are polynomials with integer coefficients, and likewise for
$z_2^{\ell_2}$. Note that since the magnitudes of $\ell_1$ and
$\ell_2$ are polynomial in $||\uu||$, so are $||A_1||$, $||B_1||$,
$||A_2||$, and $||B_2||$. The case in which $\ell_1$ or $\ell_2$ is
negative is handled similarly, with the additional use of a division
operation.

Combining everything, we obtain a sentence $\tau$ of the first-order
theory of the reals with division which is true iff
Assertion~(\ref{min-eqn}) holds. $\tau$ makes use of at most $9$ real
variables: two for each of $z_1$ and $z_2$, one for $a$, and one for
each of $\mathrm{Re}(c_1)$, $\mathrm{Im}(c_1)$, $\mathrm{Re}(c_2),
\mathrm{Im}(c_2)$. In removing divisions from $\tau$, the number of
variables potentially increases to $11$. Finally, the size of $\tau$
is polynomial in $||\uu||$. We can therefore invoke
Theorem~\ref{renegar} to conclude that Assertion~(\ref{min-eqn}) can
be decided in time polynomial in $||\uu||$.

This completes the proofs of Theorems~\ref{theorem-pos} and
\ref{theorem-upos}.

\section{Hardness at Order Six}
\label{hardness}

Diophantine approximation is an old branch of number theory concerned,
among other things, with problems related to approximating real
numbers by rationals. It is a vast and active field of research with
several hard, longstanding open problems. In this section, we present
reductions from some of these open problems to questions of Positivity
and Ultimate Positivity of integer LRS of order $6$, and \emph{a
  fortiori} of higher orders. In other words, we show that \emph{if}
Positivity or Ultimate Positivity are decidable for integer LRS of
order $6$, then certain hard open problems in Diophantine
approximation become solvable.

We survey in cursory manner some of the key definitions and facts that
are needed for our development. Results are stated largely without
proofs---comprehensive reference works include~\cite{Bak75,Sch80,RP92}.

For any real number $x$, the \defemph{Lagrange constant} (or
\defemph{homogeneous Diophantine approximation constant})
$L_{\infty}(x)$ measures the extent to which $x$ can be
`well-approximated' by rationals. It is defined as follows:
\begin{align*}
L_{\infty}(x) = \inf \smash{\Big\{} c \in \mathbb{R}: & \,
\left|x - \frac{n}{m}\right| < \frac{c}{m^2} \\
& \, \mbox{ for infinitely many $n,m \in \mathbb{Z}$} \smash{\Big\}}\, .
\end{align*}
Following Lagarias and Shallit's terminology~\cite{LS97}, we also
define the (\defemph{ho\-mo\-ge\-neous Diophantine approximation})
\defemph{type} of $x$:
\[
L(x) = \inf \left\{ c \in \mathbb{R}:
\left|x - \frac{n}{m}\right| < \frac{c}{m^2}
\mbox{ for some $n,m \in \mathbb{Z}$} \right\} .
\]

Khinchin showed in 1926 that almost all real numbers (in the
measure-theoretic sense) have Lagrange constant and type equal to
zero. Yet real numbers with non-zero Lagrange constant constitute an
uncountable class known as the \emph{badly approximable} numbers.  The
Lagrange constant and type of a real number $x$ are closely linked to
the continued fraction expansion of $x$, a fact which enabled Euler to
prove that all algebraic numbers of degree $2$ are badly approximable.

An old observation of Dirichlet shows that every real number has
Lagrange constant at most $1$. This bound was improved to $1/\sqrt{5}$
by Hurwitz in 1891, who also showed that it is achieved by the golden
ratio. Markov proved in 1879 that every transcendental real number $x$
has $L_{\infty}(x) \in [0,1/3]$. Considerable further work has been
devoted to the study of the Lagrange spectrum, which records the
possible values taken on by Lagrange constants---see,
e.g.,~\cite{CF89}. Despite this, nothing further is known about the
Lagrange constant or type of the vast majority of transcendental
numbers; for example, it is a longstanding open problem as to whether
$L_\infty(\pi)$ is $0$, $1/3$, or some value in between.

Let \[
\mathcal{A} = \{p + qi \in \mathbb{C} :
p,q \in \mathbb{Q},\, p^2 + q^2 = 1, \mbox{ and } p,q \neq 0\}
\]
be the set of points on the unit circle in the complex plane with
rational real and imaginary parts, excluding $\{1,-1,i,-i\}$. Note
that this set is dense since $\frac{1-r^2}{1+r^2} + i
\frac{2r}{1+r^2}$ always lies of the unit circle for any $r \in
\mathbb{Q}$.  Clearly, $\mathcal{A}$ consists of algebraic numbers of
degree $2$, none of which is a root of unity: indeed, the primitive
$k$th roots of unity are precisely the roots of the $k$th cyclotomic
polynomial, whose degree is $\varphi(k)$, where $\varphi$ is Euler's
totient function. Standard lower bounds on the latter imply that the
only roots of unity of degree $2$ are the $3$rd, $4$th, and $6$th
primitive roots of unity, all of which either have irrational
imaginary part or are $\pm i$. 

Write 
\[ \mathcal{T} = \left\{\frac{\arg \alpha }{2\pi} : 
\alpha \in \mathcal{A}\right\}\, .
\]
$\mathcal{T}$ is a dense subset of $(-1/2,1/2)$ consisting exclusively
  of transcendental numbers: indeed, for $t = \frac{\arg
    \alpha}{2\pi}$, we have $\alpha = e^{2\pi i t} = (-1)^{2t}$. Since
  $\alpha$ is not a root of unity, $t$ cannot be rational, and it
  follows that $t$ must be transcendental by the Gelfond-Schneider
  theorem (see, e.g,~\cite{Bak75}).

Recall that a real number $x$ is \defemph{computable} if there is an
algorithm which, given any rational $\varepsilon > 0$ as input,
returns a rational $q$ such that $|q-x| < \varepsilon$. We can now
state our main hardness results:

\begin{theorem}
\label{hard-UP}
Suppose that Ultimate Positivity is decidable for integer LRS of order
$6$. Then, for any $t \in \mathcal{T}$, $L_{\infty}(t)$ is a
computable number.
\end{theorem}

\begin{theorem}
\label{hard-P}
Suppose that Positivity is decidable for integer LRS of order
$6$. Then, for any $t \in \mathcal{T}$, $L(t)$ is a computable number.
\end{theorem}

Theorems~\ref{hard-UP} and \ref{hard-P} strongly suggest that the
decidability of Positivity and Ultimate Positivity for LRS of order 6
(and \emph{a fortiori} higher orders) are unlikely to be achievable
without major breakthroughs in analytic number theory. These theorems
also have partial converses (which are omitted in the interest of
brevity) which entail that, at least at order $6$, proofs of
\emph{undecidability} would also have substantial implications
regarding the Diophantine approximation of certain transcendental
numbers.

We now proceed with the proof of both theorems.

Choose $p + qi \in \mathcal{A}$ and $r \in \mathbb{Q}$ such that $r >
0$. Let $\theta = \arg (p+qi)$, and write
\begin{align*}
u_n = &\ r \sin n \theta - n(1 - \cos n \theta)   \\
v_n = &\ {-r} \sin n \theta - n(1 - \cos n \theta)  \, .
\end{align*}

It is not hard to see that $\uu = \langle u_n \rangle_{n=0}^{\infty}$
and $\vv = \langle v_n \rangle_{n=0}^{\infty}$ are both rational LRS
of order $6$. Indeed, writing $\lambda = p + qi$, both $\uu$ and $\vv$
are LRS with characteristic roots $1$, $\lambda$, and
$\overline{\lambda}$, each of which has multiplicity 2. The
exponential polynomial solution for $\uu$ is
\[ u_n = -n 1^n + \frac{1}{2}(n-ri)\lambda^n +
\frac{1}{2}(n+ri)\overline{\lambda}^n \, ,
\]
from which the order-$6$ recurrence relation can easily be
extracted. Note that since $\lambda$ and $\overline{\lambda}$ have
rational real and imaginary parts, by induction $u_n$ is rational for all
$n \geq 0$. Naturally, a similar exercise can be carried out for $\vv$. 

For $n \geq 0$, let 
\[
w_n = \max\{u_n,v_n\} = r|{\sin n\theta}| - n(1 - \cos n\theta) \, .
\]

Given $\varepsilon \in (0,1)$, there exists $\delta > 0$ such that, for all 
$x \in [-\delta,\delta]$, we have
\begin{gather}
\label{sin-eq}
(1-\varepsilon)|x| \leq |{\sin x}| \leq |x| \, , \mbox{ and} \\
(1-\varepsilon)\frac{x^2}{2} \leq 1 - \cos x \leq \frac{x^2}{2}
\label{cos-eq}
\end{gather}
Moreover, there exists $N \in \mathbb{N}$ with $2r/N \leq \delta$
such that, for all $x \in (-\pi,\pi]$,
\begin{equation}
\label{eq-2rN}
\mbox{if }\ 1 - \cos x < \frac{2r}{N}, \mbox{ then } |x| \leq \delta \,
.
\end{equation}

For $x \in \mathbb{R}$, recall that $[x]_{2 \pi}$ denotes the distance
from $x$ to the closest integer multiple of $2 \pi$. Let $t =
\theta/2\pi$. It is straightforward to show that
\begin{equation}
\label{L-infinity}
2 \pi L_{\infty}(t) = \liminf_{m \in \mathbb{N}} m[m(2\pi t)]_{2\pi}
= \liminf_{m \in \mathbb{N}} m[m \theta]_{2\pi}
\end{equation}
and
\begin{equation}
\label{L-type}
2 \pi L(t) = \inf_{m \in \mathbb{N}} m[m(2\pi t)]_{2\pi} 
= \inf_{m \in \mathbb{N}} m[m \theta]_{2\pi} \, .
\end{equation}

We now assert the following:
\begin{enumerate}
\item[] Claim 1: For any $m \geq N$, if $w_m > 0$, 
then $\displaystyle{m[m\theta]_{2\pi} < \frac{2r}{1-\varepsilon}}$.

\item[] Claim 2: For any $m \geq N$, if 
$m[m\theta]_{2\pi} < 2r(1-\varepsilon)$, then $w_m > 0$.
\end{enumerate}

To prove Claim~1, assume that $m \geq N$ and $w_m > 0$. Then:
\begin{alignat*}{2}
& m(1 - \cos m\theta) < r 
        \qquad \qquad \quad \ \mbox{[by definition of $w_m$]} \\
\Rightarrow \ \
& 1 - \cos m\theta < \frac{r}{m} < \frac{2r}{m} \leq \frac{2r}{N} \\
\Rightarrow \ \
& [m \theta]_{2\pi} \leq \delta 
    \qquad \qquad \qquad \qquad \qquad \qquad \quad \,
\mbox{[by~(\ref{eq-2rN})]} \\
\Rightarrow \ \
& 0 < w_m \\
& \phantom{0} \leq r[m\theta]_{2\pi} - 
           m(1-\varepsilon)\frac{([m\theta]_{2\pi})^2}{2} 
        \quad \ \mbox{[(\ref{sin-eq}), (\ref{cos-eq})]} \\
\Rightarrow \ \
& m[m\theta]_{2\pi} < \frac{2r}{1-\varepsilon} \, ,
\end{alignat*}
as required.

For Claim~2, assume that $m \geq N$ and $m[m\theta]_{2\pi} <
2r(1-\varepsilon)$. Then $[m \theta]_{2\pi} \leq 2r/N \leq \delta$,
whence 
$\displaystyle{
w_n \geq r(1-\varepsilon)[m\theta]_{2\pi} 
- m\frac{([m\theta]_{2\pi})^2}{2} 
}$ by (\ref{sin-eq}) and~(\ref{cos-eq}),
and also 
$\displaystyle{
m \frac{([m \theta]_{2\pi})^2}{2} <
r(1-\varepsilon)[m\theta]_{2\pi}
}$.\footnote{Recall that $p+qi$ is not a root of unity, and hence 
$[m\theta]_{2\pi} \neq 0$.}
Combining the last two inequalities yields $w_n > 0$ as required.

Observe that if $-\uu$ and $-\vv$ are both ultimately
positive,\footnote{Recall from Section~\ref{decidability} that
  decision procedures for Positivity and Ultimate Positivity of
  integer LRS are readily applicable to rational LRS\@.} then for all
sufficiently large $m$, we have $w_m \leq 0$, and therefore, by
Claim~2, $m[m\theta]_{2\pi} \geq 2r(1-\varepsilon)$. Since this holds
for all $\varepsilon \in (0,1)$, it follows from
Equation~(\ref{L-infinity}) that $L_{\infty}(t) \geq r/\pi$.

On the other hand, if one or both of $-\uu$ and $-\vv$ fail to be
ultimately positive, then there must be infinitely many values of $m$
such that $w_m > 0$. Claim~1 and Equation~(\ref{L-infinity}) then entail
that $L_{\infty}(t) \leq r/\pi$.

Since $r$ can be chosen arbitrarily, this establishes
Theorem~\ref{hard-UP}.

A similar procedure can be used to approximate $L(t)$. Note that
arbitrarily good upper bounds can always be guessed and, if correct,
be verified effectively, by enumerating pairs of integers until a
suitable pair is found.\footnote{Note that this requires some
  numerical analysis, which we take for granted, in order to perform
  approximations with sufficient precision.}

Suppose now that we wish to validate a purported lower bound $b <
L(t)$. Guess rational values of $r$ and $\varepsilon$ such that $2 \pi
b < 2r(1-\varepsilon) < \frac{2r}{1-\varepsilon} < 2 \pi L(t)$. Note
that one can readily compute the value of the corresponding integer
$N$ in the notation of our proof. Invoke the Positivity oracle on the
LRS $\langle -u_m \rangle_{m=N}^{\infty}$ and $\langle -v_m
\rangle_{m=N}^{\infty}$. The outcome must be that both are positive,
otherwise there would be some value of $m \geq N$ such that $w_m > 0$,
from which we would conclude via Claim~1 that $m[m\theta]_{2 \pi} <
\frac{2r}{1-\varepsilon}$, contradicting our assumption that
$\frac{2r}{1-\varepsilon} < 2\pi L(t)$.

Since both LRS are revealed to be positive, we know that for all $m
\geq N$, $w_m \leq 0$ and therefore (thanks to Claim~2) that
$m[m\theta]_{2\pi} \geq 2r(1-\varepsilon)$. It now suffices to verify
individually each value of $m \in \{0, \ldots, N-1\}$ to conclude that
$2 \pi L(t) \geq 2r(1-\varepsilon) > 2 \pi b$, as required. This completes the
proof of Theorem~\ref{hard-P}.

Let us finally remark that other hardness results, similar in both
form and spirit to Theorems~\ref{hard-UP} and \ref{hard-P}, can also
be formulated, notably via the use of techniques on
\emph{inhomogeneous} Diophantine approximation of certain
transcendental numbers.

\section{Extensions and Future Work}
\label{conclusion}

Several of the results presented in this paper have natural extensions or
generalisations, some of which we briefly mention here.

Define an LRS $\uu = \langle u_n \rangle_{n=0}^{\infty}$ to be
\defemph{strictly positive} (respectively \defemph{ultimately strictly
  positive}) if $u_n > 0$ for all $n$ (respectively for all
sufficiently large $n$). An examination of our proofs readily shows
that all our decidability and complexity results, with the exception
of the decidability and complexity of Positivity for integer LRS of
order $5$, carry over without difficulty to the analogous strict
formulation. A useful observation in this regard is that for
non-degenerate LRS, Ultimate Positivity and Ultimate Strict Positivity
agree: indeed, as can be seen from the proof of the Skolem-Mahler-Lech
theorem~\cite{BOOK}, any non-degenerate LRS is either identically zero
or has only finitely many zeros. Let us also mention that our
Diophantine-approximation hardness results are easily seen to carry
over \emph{mutatis mutandis} to Strict Positivity and Ultimate Strict
Positivity.

All our decidability results also carry over to LRS over real
algebraic numbers, as can readily be seen by examining the relevant
proofs. Our complexity upper bounds, however, are more delicate, and
it is an open question whether they continue to hold in the algebraic
setting. Hardness results, on the other hand, obviously carry over to
the more general algebraic world.

It seems likely that the techniques developed in this paper could be
usefully deployed to tackle other natural decision problems for linear
recurrence sequences, such as divergence to infinity, reachability and
ultimate reachability of semi-linear sets, etc. In turn such decision
procedures---or corresponding hardness results---may find applications
in some of the areas mentioned in the Introduction, such as the
analysis of termination of linear programs or the behaviour of
discrete linear dynamical systems. More ambitiously, in the spirit of
synthesis, one could seek to explore computational problems for
\emph{parametric} LRS, where the aim is to characterise ranges for the
parameters guaranteeing certain behaviours, etc.

Another interesting question concerns the complexity of Positivity at
low orders. Recall that the $\mathrm{PosSLP}$ oracle used in our main
decision procedure is invoked to check whether the quantity $\vec{v}^T
M^n \vec{w}$ is strictly negative, where $M$ is a $k \times k$ matrix
of integers, $\vec{v}$ and $\vec{w}$ are $k$-dimensional integer
vectors, and $n$ is encoded in binary. It is conceivable---especially
for small fixed $k$, as in the situation at hand---that the complexity
of this problem is significantly lower than that of
$\mathrm{PosSLP}$. See~\cite{HKR10} for initial progress on related
questions.

Finally, the various discrete problems discussed in the present paper
also have natural counterparts in a continuous
setting. See~\cite{BDJB10}, for example, which studies the Skolem and
Positivity Problems over continuous time using similar tools. This
remains a largely unexplored research landscape.

\bibliography{posbib2.bib}  

\begin{thebibliography}{10}

\bibitem{AAG12}
M.~Agrawal, S.~Akshay, B.~Genest, and P.~S. Thiagarajan.
\newblock Approximate verification of the symbolic dynamics of {M}arkov chains.
\newblock In {\em Proc. Symp. on Logic in Comp. Sci. (LICS)}. IEEE, 2012.

\bibitem{ABK09}
E.~Allender, P.~B\"{u}rgisser, J.~Kjeldgaard-Pedersen, and P.~B. Miltersen.
\newblock On the complexity of numerical analysis.
\newblock {\em SIAM J. Comput.}, 38(5), 2009.

\bibitem{Bak75}
A.~Baker.
\newblock {\em Transcendental Number Theory}.
\newblock Cambridge University Press, 1975.

\bibitem{BW93}
A.~Baker and G.~W\"{u}stholz.
\newblock Logarithmic forms and group varieties.
\newblock {\em Jour. Reine Angew. Math.}, 442, 1993.

\bibitem{BPR06}
S.~Basu, R.~Pollack, and M.-F. Roy.
\newblock {\em Algorithms in Real Algebraic Geometry}.
\newblock Springer, 2nd edition, 2006.

\bibitem{Bau70}
W.~J. Baumol.
\newblock {\em Economic Dynamics}.
\newblock Prentice Hall, 3rd edition, 1970.

\bibitem{BRS06}
D.~Beauquier, A.~M. Rabinovich, and A.~Slissenko.
\newblock A logic of probability with decidable model checking.
\newblock {\em J. Log. Comput.}, 16(4), 2006.

\bibitem{BG07}
J.~P. Bell and S.~Gerhold.
\newblock On the positivity set of a linear recurrence.
\newblock {\em Israel Jour. Math.}, 57, 2007.

\bibitem{BDJB10}
P.~C. Bell, J.-C. Delvenne, R.~M. Jungers, and V.~D. Blondel.
\newblock The continuous {S}kolem-{P}isot problem.
\newblock {\em Theor. Comput. Sci.}, 411(40-42), 2010.

\bibitem{Ben13}
A.~M. Ben-Amram.
\newblock Mortality of iterated piecewise affine functions over the integers:
  Decidability and complexity.
\newblock In {\em Proc. Intern. Symp. on Theoret. Aspects of Comp.
  Sci.(STACS)}, volume~20 of {\em LIPIcs}. Schloss Dagstuhl -- Leibniz-Zentrum
  f\"ur Informatik, 2013.

\bibitem{BGM12}
A.~M. Ben-Amram, S.~Genaim, and A.~N. Masud.
\newblock On the termination of integer loops.
\newblock {\em ACM Trans. Program. Lang. Syst.}, 34(4), 2012.

\bibitem{BM76}
J.~Berstel and M.~Mignotte.
\newblock Deux propri\'et\'es d\'ecidables des suites r\'ecurrentes
  lin\'eaires.
\newblock {\em Bull. Soc. Math. France}, 104, 1976.

\bibitem{BJK05}
V.~D. Blondel, E.~Jeandel, P.~Koiran, and N.~Portier.
\newblock Decidable and undecidable problems about quantum automata.
\newblock {\em SIAM J. Comput.}, 34(6), 2005.

\bibitem{BP02}
V.~D. Blondel and N.~Portier.
\newblock The presence of a zero in an integer linear recurrent sequence is
  {NP}-hard to decide.
\newblock {\em Linear Algebra and Its Applications}, 351--352, 2002.

\bibitem{Bou66}
N.~Bourbaki.
\newblock {\em Elements of Mathematics: General Topology (Part 2)}.
\newblock Addison-Wesley, 1966.

\bibitem{BIK12}
M.~Bozga, R.~Iosif, and Filip Konecn{\'y}.
\newblock Deciding conditional termination.
\newblock In {\em Proc. Intern. Conf. on Tools and Algorithms for the
  Construction and Analysis of Systems (TACAS)}, volume 7214 of {\em LNCS}.
  Springer, 2012.

\bibitem{Bra06}
M.~Braverman.
\newblock Termination of integer linear programs.
\newblock In {\em Proc. Intern. Conf. on Computer Aided Verification (CAV)},
  volume 4144 of {\em LNCS}. Springer, 2006.

\bibitem{BW81}
J.~R. Burke and W.~A. Webb.
\newblock Asymptotic behavior of linear recurrences.
\newblock {\em Fib. Quart.}, 19(4), 1981.

\bibitem{CLZ00}
J.-Y. Cai, R.~J. Lipton, and Y.~Zalcstein.
\newblock The complexity of the {A B C} problem.
\newblock {\em SIAM J. Comput.}, 29(6), 2000.

\bibitem{COW13}
V.~Chonev, J.~Ouaknine, and J.~Worrell.
\newblock The {O}rbit {P}roblem in higher dimensions.
\newblock In {\em Proc. Symp. on the Theory of Computing (STOC)}. ACM, 2013.

\bibitem{Coh93}
H.~Cohen.
\newblock {\em A Course in Computational Algebraic Number Theory}.
\newblock Springer-Verlag, 1993.

\bibitem{Col75}
G.~E. Collins.
\newblock Quantifier elimination for real closed fields by cylindrical
  algebraic decomposition.
\newblock In {\em Proc. 2nd GI Conf. Automata Theory and Formal Languages}.
  Springer-Verlag, 1975.

\bibitem{CPR11}
B.~Cook, A.~Podelski, and A.~Rybalchenko.
\newblock Proving program termination.
\newblock {\em Commun. ACM}, 54(5), 2011.

\bibitem{CF89}
T.~W. Cusick and M.~E. Flahive.
\newblock {\em The {M}arkoff and {L}agrange Spectra}.
\newblock American Mathematical Society, 1989.

\bibitem{DJK05}
H.~Derksen, E.~Jeandel, and P.~Koiran.
\newblock Quantum automata and algebraic groups.
\newblock {\em J. Symb. Comput.}, 39(3-4), 2005.

\bibitem{BOOK}
G.~Everest, A.~van~der Poorten, I.~Shparlinski, and T.~Ward.
\newblock {\em Recurrence Sequences}.
\newblock American Mathematical Society, 2003.

\bibitem{Ge93}
G.~Ge.
\newblock {\em Algorithms Related to Multiplicative Representations of
  Algebraic Numbers}.
\newblock PhD thesis, U.C. Berkeley, 1993.

\bibitem{GL91}
I.~Gyori and G.~Ladas.
\newblock {\em Oscillation Theory of Delay Differential Equations}.
\newblock Oxford Mathematical Monographs. Oxford University Press, 1991.

\bibitem{HHH06}
V.~Halava, T.~Harju, and M.~Hirvensalo.
\newblock Positivity of second order linear recurrent sequences.
\newblock {\em Discrete Applied Mathematics}, 154(3), 2006.

\bibitem{TUCS05}
V.~Halava, T.~Harju, M.~Hirvensalo, and J.~Karhum\"aki.
\newblock Skolem's problem --- on the border between decidability and
  undecidability.
\newblock Technical Report 683, Turku Centre for Computer Science, 2005.

\bibitem{HKR10}
M.~Hirvensalo, J.~Karhum{\"a}ki, and A.~Rabinovich.
\newblock Computing partial information out of intractable: Powers of algebraic
  numbers as an example.
\newblock {\em Jour. Number Theory}, 130, 2010.

\bibitem{KL86}
R.~Kannan and R.~J. Lipton.
\newblock Polynomial-time algorithm for the orbit problem.
\newblock {\em Jour. ACM}, 33(4), 1986.

\bibitem{LS97}
J.~C. Lagarias and J.~O. Shallit.
\newblock Linear fractional transformations of continued fractions with bounded
  partial quotients.
\newblock {\em Journal de Th\'eorie des Nombres de Bordeaux}, 9, 1997.

\bibitem{Lao13}
V.~Laohakosol.
\newblock Personal communication, July 2013.

\bibitem{LT09}
V.~Laohakosol and P.~Tangsupphathawat.
\newblock Positivity of third order linear recurrence sequences.
\newblock {\em Discrete Applied Mathematics}, 157(15), 2009.

\bibitem{LR76}
A.~Lindenmayer and G.~Rozenberg, editors.
\newblock {\em Automata, Languages, Development}. North-Holland, 1976.

\bibitem{Lip09}
R.~J. Lipton.
\newblock Mathematical embarrassments.
\newblock \emph{Blog entry}, December 2009.
\newblock \texttt{http://rjlipton.wordpress.com/2009/12/26/
  mathematical-embarrassments/}.

\bibitem{Liu10}
L.~L. Liu.
\newblock Positivity of three-term recurrence sequences.
\newblock {\em Electr. J. Comb.}, 17(1), 2010.

\bibitem{Mas88}
D.~W. Masser.
\newblock Linear relations on algebraic groups.
\newblock In {\em New Advances in Transcendence Theory}. Cambridge University
  Press, 1988.

\bibitem{Mig82}
M.~Mignotte.
\newblock Some useful bounds.
\newblock In {\em Computer Algebra}, 1982.

\bibitem{MST84}
M.~Mignotte, T.~N. Shorey, and R.~Tijdeman.
\newblock The distance between terms of an algebraic recurrence sequence.
\newblock {\em Journal f\"ur die reine und angewandte Mathematik}, 349, 1984.

\bibitem{NS90}
K.~Nagasaka and J.-S. Shiue.
\newblock Asymptotic positiveness of linear recurrence sequences.
\newblock {\em Fib. Quart.}, 28(4), 1990.

\bibitem{OW12}
J.~Ouaknine and J.~Worrell.
\newblock Decision problems for linear recurrence sequences.
\newblock In {\em Proc. Intern. Workshop on Reachability Problems (RP)}, volume
  7550 of {\em LNCS}. Springer, 2012.

\bibitem{Pan97}
V.~Pan.
\newblock Optimal and nearly optimal algorithms for approximating polynomial
  zeros.
\newblock {\em Computers {\&} Mathematics with Applications}, 31(12), 1996.

\bibitem{PR04}
A.~Podelski and A.~Rybalchenko.
\newblock A complete method for the synthesis of linear ranking functions.
\newblock In {\em Proc. Intern. Conf. on Verif., Model Checking, and Abstract
  Interpretation (VMCAI)}, volume 2937 of {\em LNCS}. Springer, 2004.

\bibitem{Ren92}
J.~Renegar.
\newblock On the computational complexity and geometry of the first-order
  theory of the reals. {Part I}: {I}ntroduction. {P}reliminaries. {T}he
  geometry of semi-algebraic sets. {T}he decision problem for the existential
  theory of the reals.
\newblock {\em J. Symb. Comp.}, 1992.

\bibitem{RP92}
A.~M. Rockett and P.~Sz\"usz.
\newblock {\em Continued Fractions}.
\newblock World Scientific, 1992.

\bibitem{RS94}
G.~Rozenberg and A.~Salomaa.
\newblock {\em Cornerstones of Undecidability}.
\newblock Prentice Hall, 1994.

\bibitem{Sal76}
A.~Salomaa.
\newblock Growth functions of {L}indenmayer systems: Some new approaches.
\newblock In A.~Lindenmayer and G.~Rozenberg, editors, {\em Automata,
  Languages, Development}. North-Holland, 1976.

\bibitem{Sch80}
W.~M. Schmidt.
\newblock Diophantine approximation.
\newblock In {\em Lect. Notes. in Math.}, volume 785, 1980.

\bibitem{Soi76}
M.~Soittola.
\newblock On {D0L} synthesis problem.
\newblock In A.~Lindenmayer and G.~Rozenberg, editors, {\em Automata,
  Languages, Development}. North-Holland, 1976.

\bibitem{TPL12}
P.~Tangsupphathawat, N.~Punnim, and V.~Laohakosol.
\newblock The positivity problem for fourth order linear recurrence sequences
  is decidable.
\newblock {\em Colloq. Math.}, 128(1), 2012.

\bibitem{Tao07}
T.~Tao.
\newblock Open question: effective {S}kolem-{M}ahler-{L}ech theorem.
\newblock \emph{Blog entry}, May 2007.
\newblock \texttt{http://terrytao.wordpress.com/2007/05/25/open-
  question-effective-skolem-mahler-lech-theorem/}.

\bibitem{TV11}
S.~P. Tarasov and M.~N. Vyalyi.
\newblock Orbits of linear maps and regular languages.
\newblock In {\em Proc. Intern. Comp. Sci. Symp. in Russia (CSR)}, volume 6651
  of {\em LNCS}. Springer, 2011.

\bibitem{Tar51}
A.~Tarski.
\newblock {\em A Decision Method for Elementary Algebra and Geometry}.
\newblock University of California Press, 1951.

\bibitem{Tiw04}
A.~Tiwari.
\newblock Termination of linear programs.
\newblock In {\em Proc. Intern. Conf. on Comp. Aided Verif. (CAV)}, volume 3114
  of {\em LNCS}. Springer, 2004.

\bibitem{Ver85}
N.~K. Vereshchagin.
\newblock The problem of appearance of a zero in a linear recurrence sequence
  (in {R}ussian).
\newblock {\em Mat. Zametki}, 38(2), 1985.

\bibitem{YLN95}
K.~Yokoyama, Z.~Li, and I.~Nemes.
\newblock Finding roots of unity among quotients of the roots of an integral
  polynomial.
\newblock In {\em Proc. Intern. Symp. on Symb. and Algebraic Comp.}, 1995.

\end{thebibliography}

\end{document}